\documentclass[preprint]{elsarticle}

\journal{High Energy Density Physics}

\usepackage{graphicx}

\begin{document}

\begin{frontmatter}
\title{Creation of Magnetized Jet Using a Ring of Laser Beams}
\author[rice]{Wen Fu}
  \ead{Wen.Fu@rice.edu}
\author[rice]{Edison P. Liang}
\address[rice]{Department of Physics and Astronomy, Rice University, Houston, TX 77005, USA}
\author[chicago]{Petros Tzeferacos}
\author[chicago]{Donald Q. Lamb}
\address[chicago]{Flash Center for Computational Science, University of Chicago, Chicago, IL 60637, USA}

\begin{abstract}
We propose a new way of generating magnetized supersonic jets using a ring laser to irradiate a flat surface target. Using 2D FLASH code simulations which include the Biermann Battery term, we demonstrate that strong toroidal fields can be generated and sustained downstream in the collimated jet outflow far from the target surface. The field strength can be controlled by varying the ring laser separation, thereby providing a versatile laboratory platform for studying the effects of magnetic field in a variety of astrophysical settings.
\end{abstract}


\end{frontmatter}

\section{Introduction}
Using FLASH code \cite{Fryxell00} simulations, we have recently demonstrated that a bundle of laser beams of given individual intensity, duration and focal spot size, produces a supersonic jet of higher density, temperature, velocity and collimation, if the beams are focused to form a circular ring pattern at the target instead of a single focal spot \cite{Fu13}. The increased density, temperature and velocity provide a more versatile platform for a variety of laboratory astrophysics experiments \cite{Ross12, Park12, Kugland12, Grosskopf13} with larger dynamic range. In this paper we investigate the effects of the ring pattern laser configuration on magnetic field generation, using a new version of the FLASH code, which includes the Biermann Battery (BB) or $(\nabla P_e \times \nabla n_e)$ term \cite{Biermann50, Krall86} for magnetic field creation, where $P_e$ and $n_e$ are electron pressure and density, respectively. Because of the 2D nature of our simulations, only the azimuthal $B_{\phi}$ filed is created. However, it is well known that for quasi-axisymmetric laser-driven blowoffs, $B_{\phi}$ is the dominant field component generated by the BB effect, so that our 2D results should be valid to first order. In follow-up simulations, we will use the 3D FLASH code (without the BB term) to check that the $(r,\,z)$ components of the BB term remain small compared to the $\phi$-component throughout, so that our 2D approximation if self-consistent. Further down the road, we will explore inclusion of the $(r,\,z)$ components of the BB term into fully 3D FLASH simulations to study various non-axisymmetric effects. 

In the current 2D $(r,\,z)$ version of the FLASH code, the magnetic field evolution terms included are \cite{Krall86}:
\begin{equation}
\frac{\partial \mathbf{B}}{\partial t}=\nabla \times (u\times \mathbf{B})-c\nabla\times(\eta \mathbf{j})+c\frac{\nabla P_e \times \nabla n_e}{en_e^2}.
\end{equation}
In the limit of low plasma resistivity $\eta$ (e.g. low density, high temperature plasma), the last term (so called ``Biermann battery'' term) dominates field creation. In 2D, $P_e$, $n_e$ are functions of only $(r,\,z)$. Hence the $\nabla P_e \times \nabla n_e$ term has only a $\phi$-component. The code also adopts a shock detection scheme to suppress artificial creation of unphysical BB fields at the shock front. Laser absorption is computed using ray-tracing in the geometric optics approximation.

\begin{figure}
\begin{center}
\includegraphics[width=0.9\textwidth]{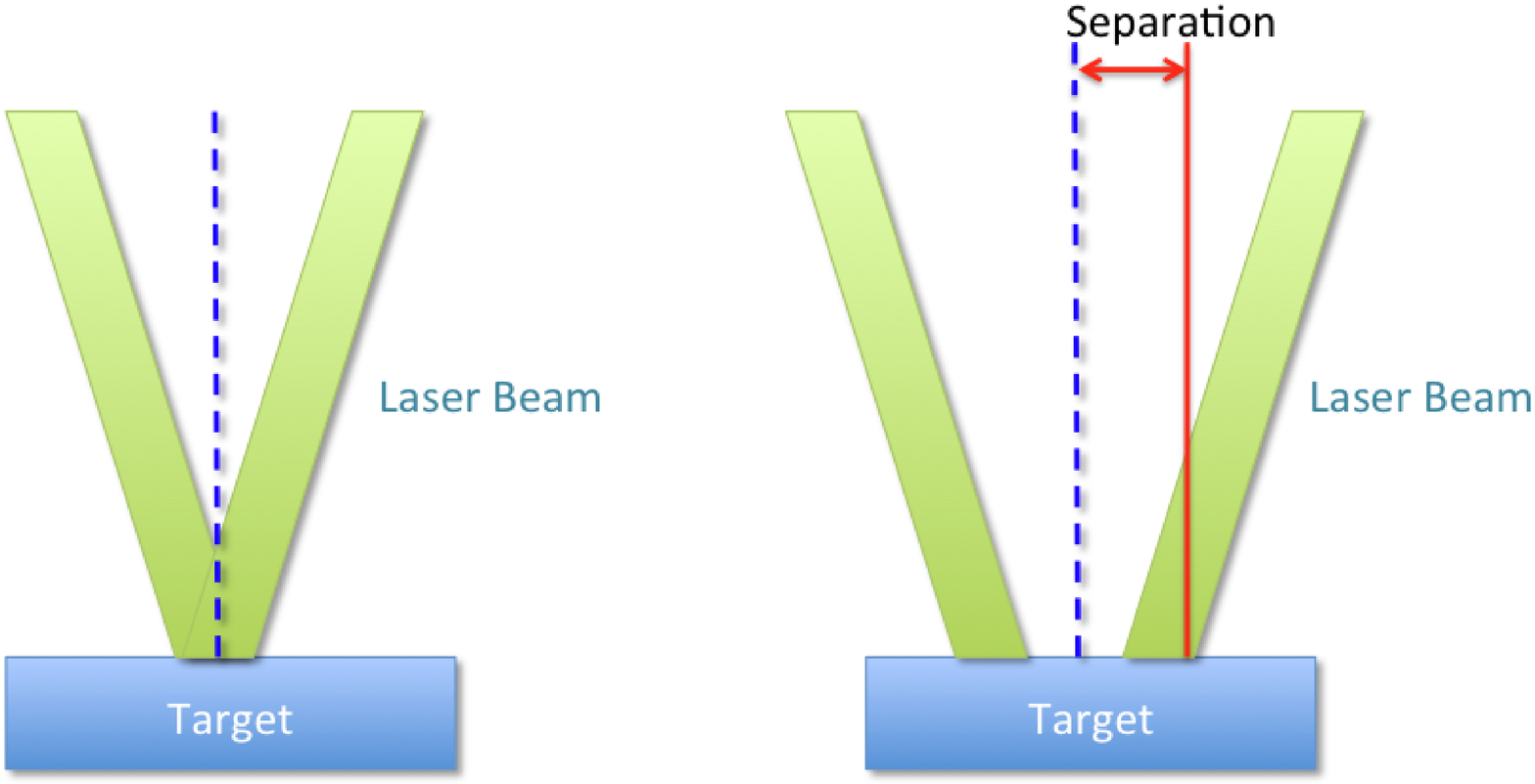}
\caption{Sketch illustrating the geometry used in our FLASH simulations}
\end{center}
\label{fig:fig1} 
\end{figure}

\section{Numerical Simulations}
We use a $(r,\,z)$ grid with $(512\times 2048)$ zones. The electron-ion plasma has zero initial magnetic field. The laser target is modeled as a $2\,mm$ wide by $0.5\,mm$ thick CH foil. We simulate the laser-driven blowoffs using laser parameters equivalent to 10 Omega laser beams of $250\,\mu m$ focal spot diameter. The total laser energy equals $5\,kJ$ in a $1\,ns$ square pulse. The laser incident angle is $30^{\circ}$ from target normal. 

\begin{figure}
\begin{center}
\includegraphics[width=1.0\textwidth]{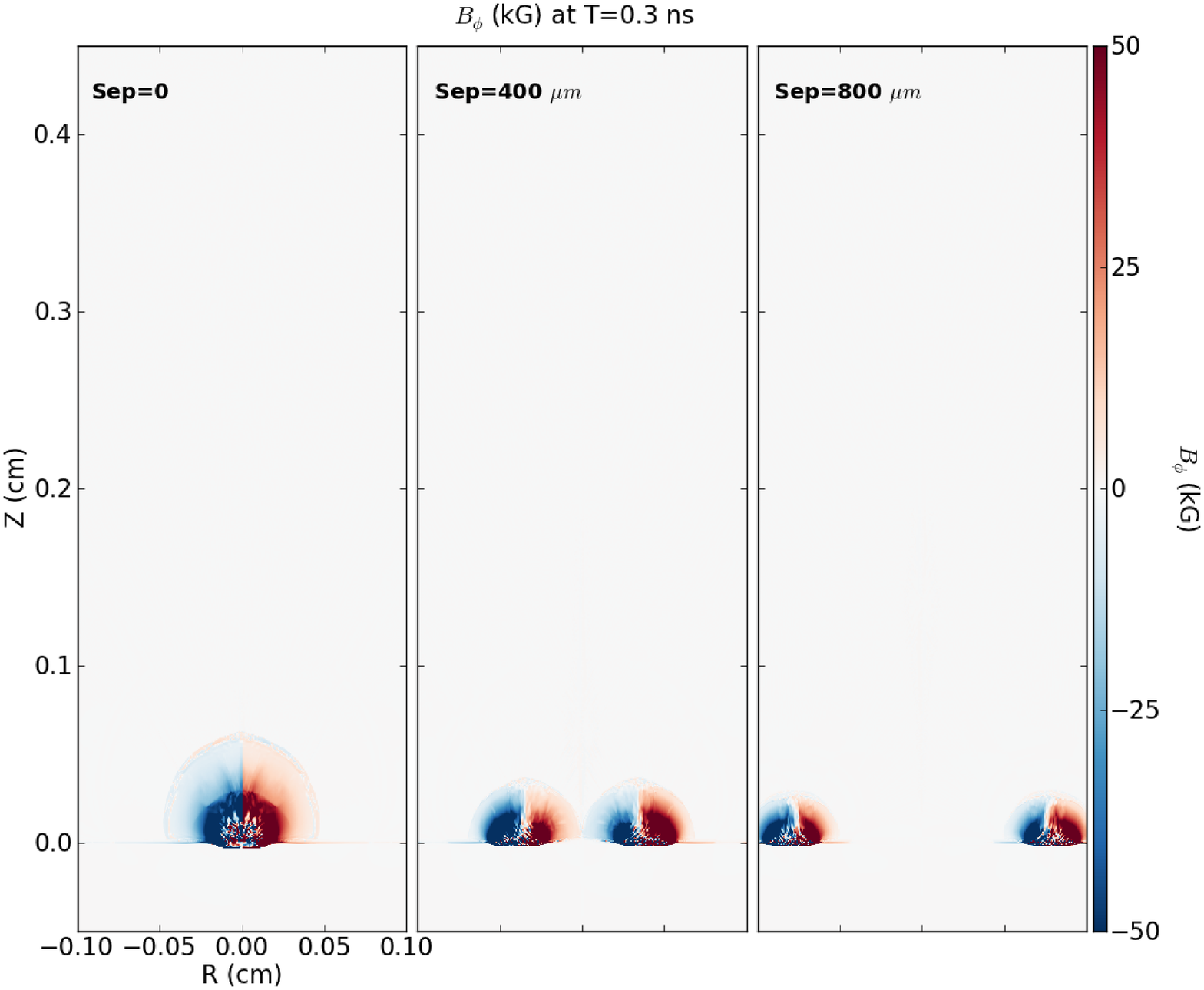}
\caption{$B_{\phi}$ profiles at t=0.3 ns for the three runs with different ring laser separations. Color scales denote field strength into and out of the plane. The toroidal fields inside and outside of the ring have opposite polarity.}
\end{center}
\label{fig:fig2} 
\end{figure}

\begin{figure}
\begin{center}
\includegraphics[width=1.0\textwidth]{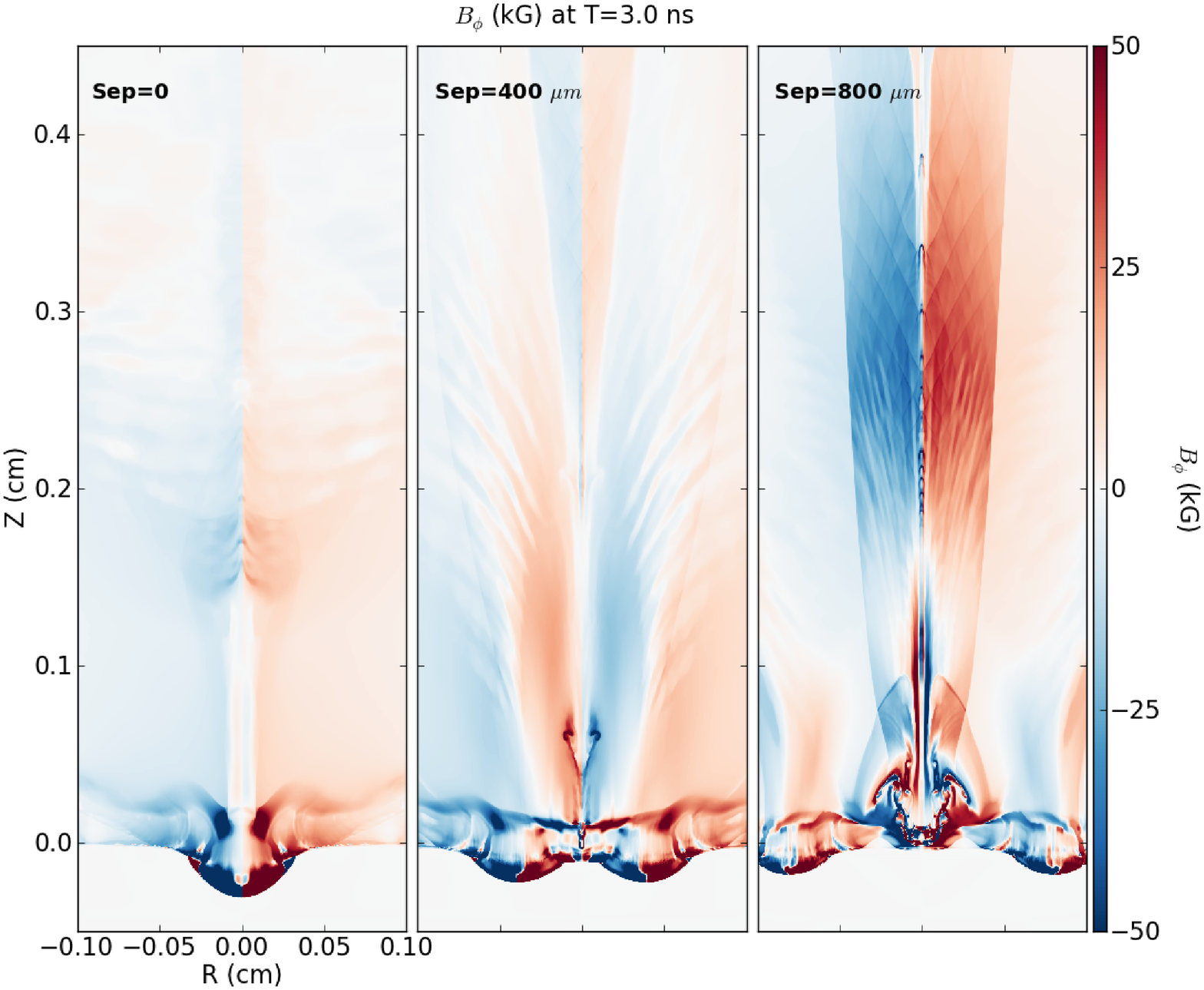}
\caption{Same as Figure \ref{fig:fig2} at t=3.0 ns. The field for the 800 $\mu m$ separation case is much stronger far from the target, due to sustained BB effects along the axis, exceeding 20 kG even at z=0.3 cm. In contrast, the field for the zero separation case is negligible by this time except near the target surface.}
\end{center}
\label{fig:fig3} 
\end{figure}

\begin{figure}
\begin{center}
$
\begin{array}{cc}
\includegraphics[width=1.0\textwidth]{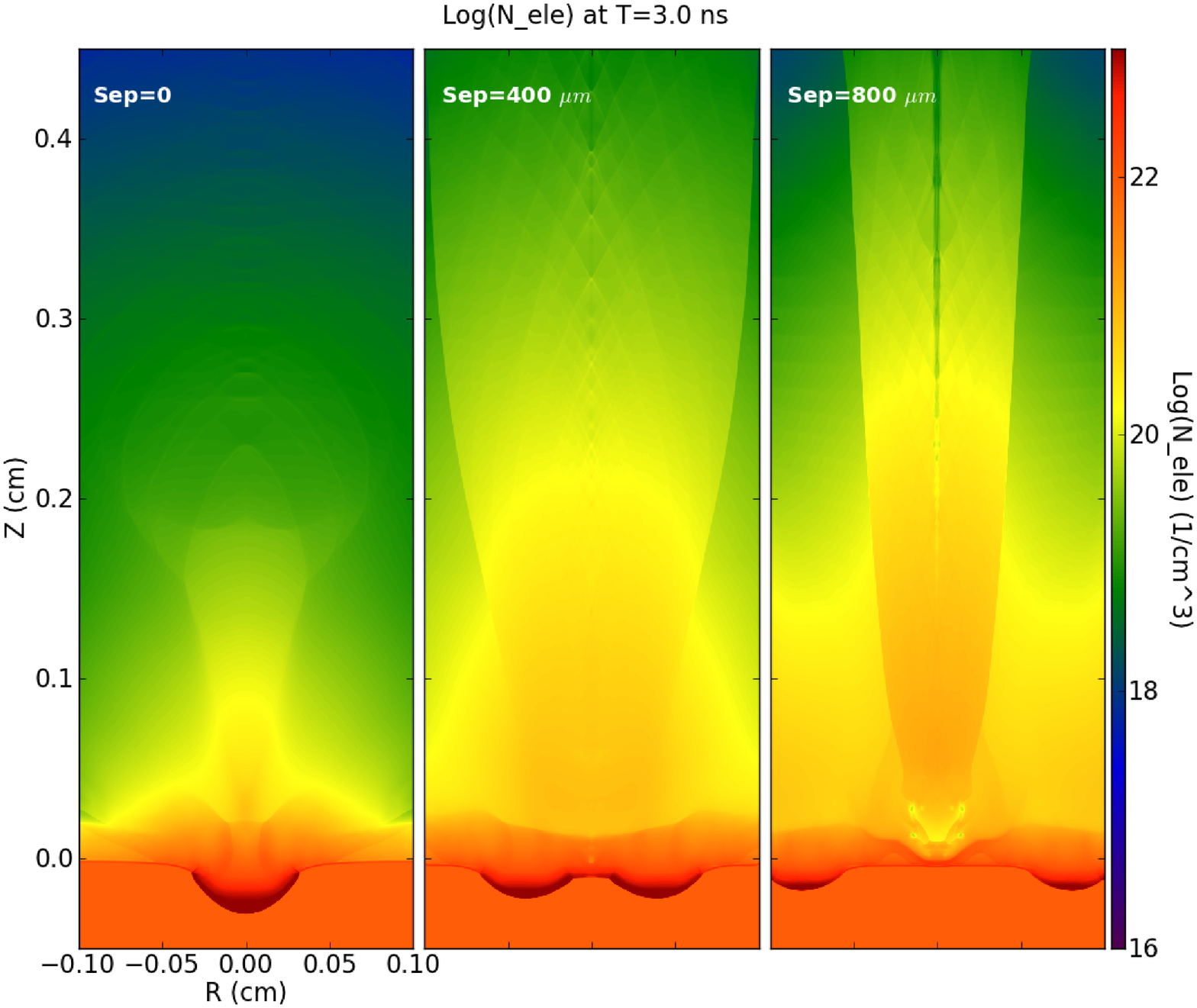} \\
\includegraphics[width=1.0\textwidth]{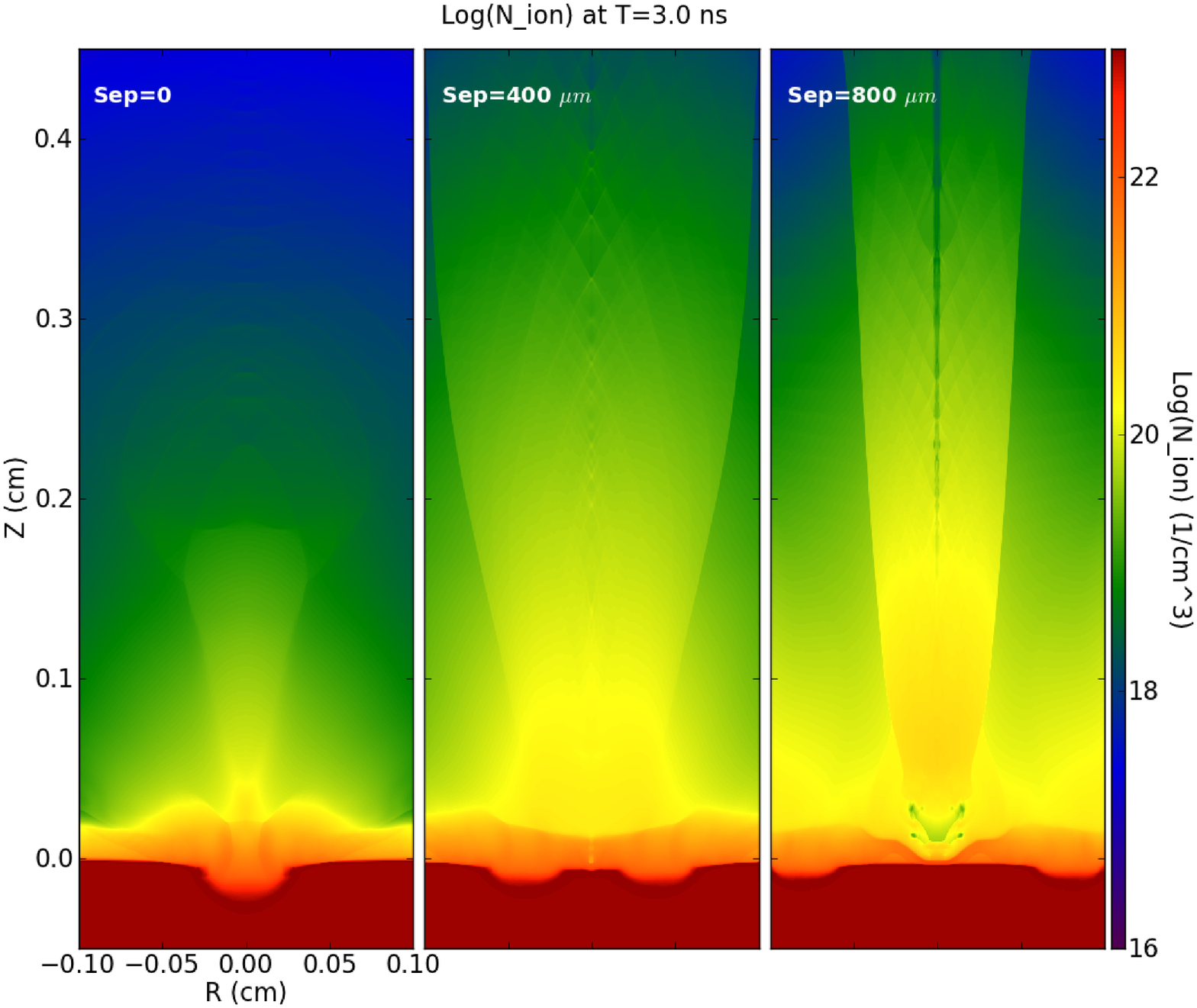} 
\end{array}
$
\caption{Electron (top)and ion (bottom) density profiles at t=3 ns showing the strong collimation and density enhancement along the axis as the ring laser separation is increased.}
\end{center}
\label{fig:fig4}
\end{figure}

\begin{figure}
\begin{center}
$
\begin{array}{cc}
\includegraphics[width=1.0\textwidth]{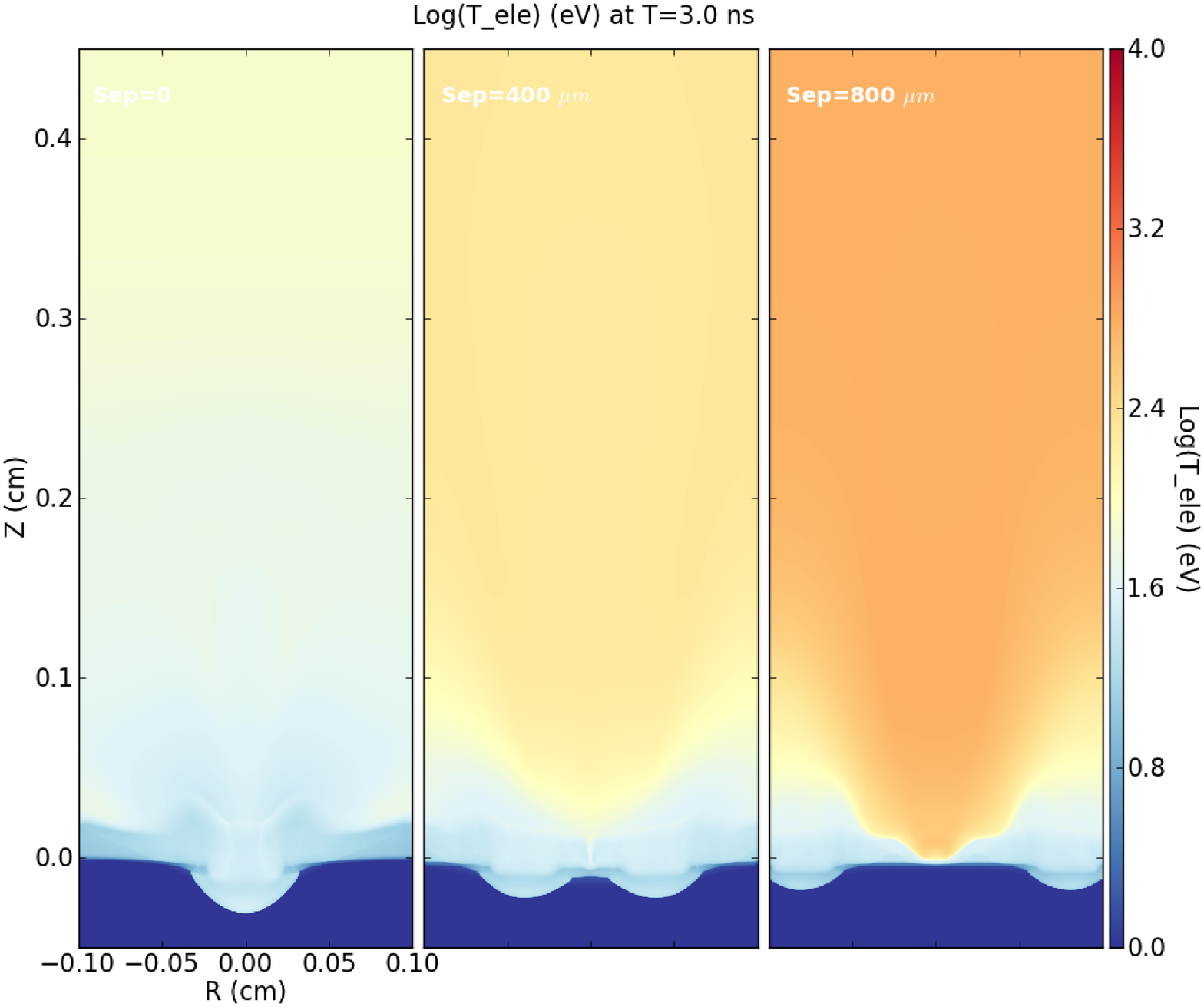} \\
\includegraphics[width=1.0\textwidth]{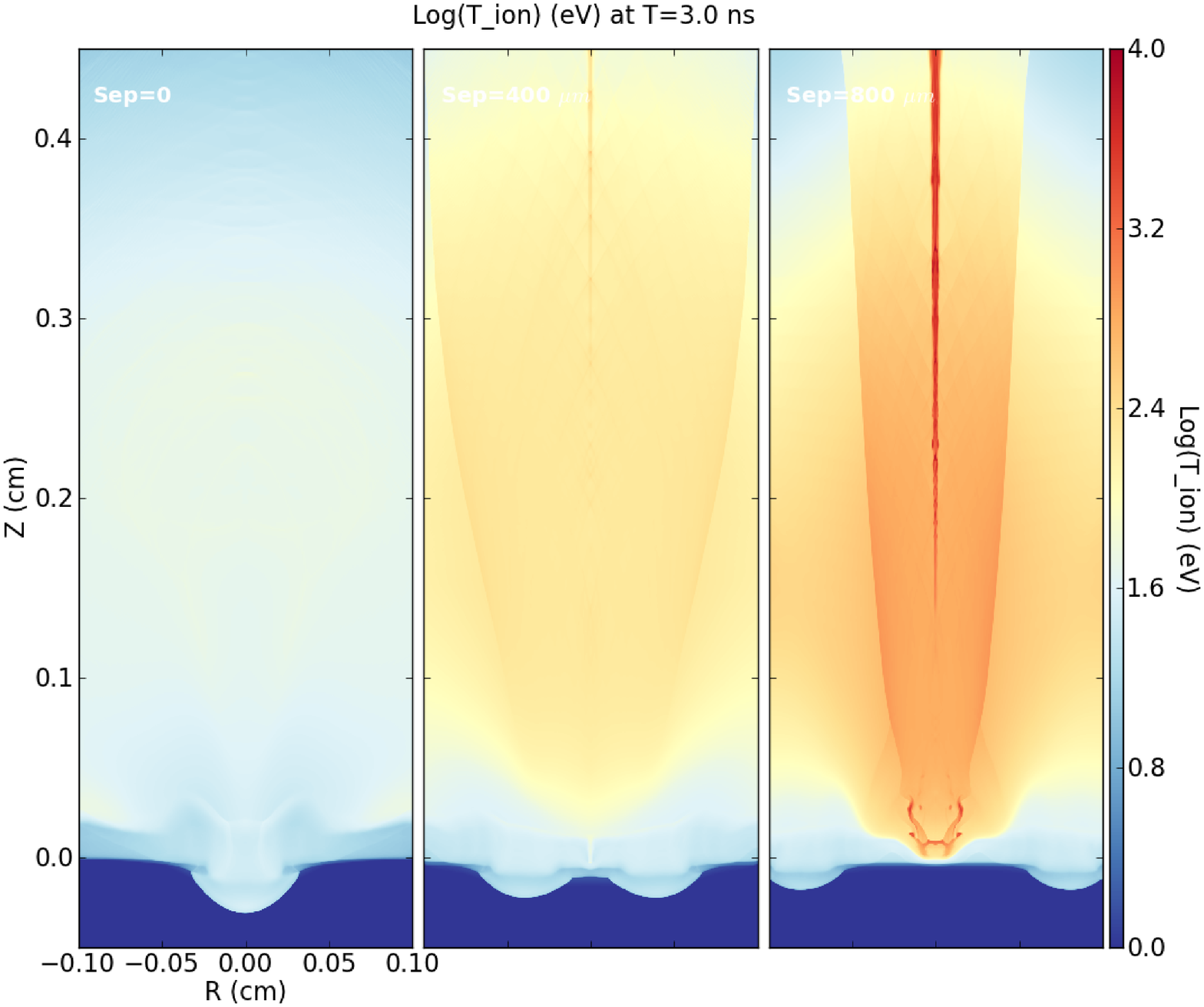} 
\end{array}
$
\caption{Electron (top)and ion (bottom) temperature profiles at t=3 ns showing enhanced heating along the axis as the ring laser separation is increased.}
\end{center}
\label{fig:fig5}
\end{figure}

Figure \ref{fig:fig1} illustrates the geometry setup of the different runs. ``Separation'' is here defined to be the distance from the midpoint of the circular ring to the target center. Figure \ref{fig:fig2} compares the BB-generated field at 0.3 ns, for three different runs: single focal spot (left), 400 $\mu m$ separation (middle), and 800 $\mu m$ separation (right). The maximum BB-generated field at this early time is around 50 kG. While the field for the single-spot laser blowoff has a simple toroidal geometry, the ring-laser blowoff creates toroidal B fields of opposite polarity on the inside and outside of the ring, as expected. Figure \ref{fig:fig3} compares the B field at 3 ns, after the laser has turned off. By this time the blowoff of the zero separation case has diverged to the point that the density and pressure gradients are too weak to sustain significant magnetic field except near the target surface. In contrast, the 800 $\mu m$ separation case shows strong toroidal field far downstream along the axis due to the compression and heating of the convergent flow from the ring towards the axis. Even at a distance of z=0.3 cm, the BB-generated field exceeds 20 kG, which would be useful for a variety of magnetized jet experiments. At the jet base, the field is more complicated due to multiple shock formation from the plasma collisions, with multiple polarity reversals due to idealization of the MHD approximation. In a real physical experiment, finite resistivity, either due to collisions or anomalous effects, will lead to reconnection and decay of the magnetic field near the base. Figure \ref{fig:fig4} compares the electron and ion densities at 3 ns, showing strong collimation and density enhancement for the ring cases. Figure \ref{fig:fig5} compares the electron and ion temperatures at 3 ns, which mirror the density enhancement. These increased density and temperature gradients are responsible for sustaining the BB-generated field downstream far from the targe surface (Figure \ref{fig:fig3}).

\textbf{Magnetic field generation via the Biermann battery term has been investigated before in various laser-plasma experiments \cite{Nishiguchi84, Willingale10a, Willingale10b}. In these studies, hot electron transport (e.g. Nernst effect leading to electron advection of magnetic fields) played an important role. However, hot electron transport is not turned on in our simulations. We are mainly interested in outflow conditions at distances $>$ 1 mm from laser target and times $>$ 1 ns. Hence transient effects at early times ($<$ ns) and near the target surface ($<$ 1 mm), which can be very complicated, will not be modeled in details as a first approximation. The FLASH code does have the option of incorporating electron transport using both the Branginski conductivity as well as flux limiters and we plan to include these in future 3D simulations to study the early evolutions of the field. Similarly fields generated by other collisionless kinetic processes such as hot electron current loops (the ``fountain effect'') and ponderomotive effect: $\nabla n_{e} \times \nabla I$ (I being laser intensity) \cite{Gibbson05} are not included in an MHD code such as FLASH. However, we believe these transport and collisionless effects primarily generate fields near the surface which are rapidly diluted volumetrically by the expanding flow and decay due to finite diffusivity. Hence they do not contribute to significant fields much beyond a distance larger than the focus spot size, i.e. a few hundred microns.}

\section{Discussion}
In our previous paper \cite{Fu13} we discussed the collisionality of the laser-driven plasma outflow, comparing the case of zero laser separation with those of 800 $\mu m$ separation. Those results, however, did not take into account the effects of BB-generated magnetic fields. Hence it is interesting to consider the effects of the BB-generated magnetic fields on the plasma physics. In the following we consider the 800 $mu m$ separation case at t=3 ns, focusing on the spatial region located $\sim$ 3 $mm$ from the target surface as a reference point. There we have an average B field of $\sim$ 20 kG, electron density $\sim$ $10^{20}$ /cc, ion density $\sim$ $10^{19}$ /cc, electron temperature $\sim$ ion temperature $\sim$ keV. For these parameters, the magnetic energy is tiny compared to the thermal energy ($E_B/E_{th}\sim 10^{-4}$). So the field has little effect on the plasma dynamics. On the other hand, the field can have important effects on the collisionality of the plasma. The electron gyroradius at this temperature and field strength is $\sim$ 38 $\mu m$, less than twice the ion skin depth $c/\omega_{pi} \sim 23 \,\mu m$. For comparison, the characteristic electrostatic and eletromagnetic instability length scales are $\sim$ 1-2 $mm$, and the Coulomb mean free path $\sim $ $cm$ (see Figure 4 of \cite{Fu13}). Hence we conclude that electron gyro-motions in the BB-generated field, together with ion plasma instabilities, will play major roles in the effective collisonality of the plasma. Hence anomalous dissipation processes will need to take into account the BB-generated magnetic field. This was not the case with the plasma outflow driven by a single laser spot (zero separation, Figure \ref{fig:fig3}). In collisionless shock experiments using two-head on colliding jets \cite{Ross12, Park12, Kugland12, Grosskopf13}, Weibel instability is invoked to generate filamentary B-fields, to mediate the shock formation and dissipation processes if the colliding plasmas are unmagnetized. However, if the colliding jets are already magnetized with perpendicular fields up to $\sim$ 20 kG, the shocks will behave very differently from pure Weibel shocks. By changing the separation of the ring laser, we can control the desired field strength of the jet, thereby giving us a new versatile laboratory platform to study the effects of magnetic field in a variety of astrophysical phenomena.

\section*{Acknowledgments}
At Rice University, this work was supported by NSF AST 1313129 and LANL Contract No. 257712-1. At the University of Chicago, this work was supported in part by the U.S. DOE through FWP 57789 under contract DE-AC02-06CH11357 to ANL. The FLASH code used in this work was developed in part by the U. S. DOE NNSA ASC- and NSF-supported Flash Center for Computational Science at the University of Chicago. Computing resource at Rice was provided by the Cyberinfrastructure for Computational Research funded by NSF under Grant CNS-0821727.


\begin{thebibliography}{}
\bibitem[Biermann (1950)]{Biermann50}
Biermann, L. 1950, Zs. Naturforsch. A., 5, 65

\bibitem[Fryxell et al. (2000)]{Fryxell00}
Fryxell, B., Olson, K., Ricker, P., et al. 2000, ApJS, 131, 273

\bibitem[Fu et al. (2013)]{Fu13}
Fu, W., Liang, E., Fatenejad, M., et al. 2013, High Energy Density Physics, 9, 336

\bibitem[Grosskopf et al. (2013)]{Grosskopf13}
Grosskopf, M, R. P. Drake, Kuranz, C. C., et al. 2013, High Energy Density Physics, 9, 192

\bibitem[Krall \& Trivelpiece (1986)]{Krall86}
Krall, N. \& Trivelpiece, A. 1986, Principles of Plasma Physics, San Francisco Press

\bibitem[Nishiguchi et al. (1984)]{Nishiguchi84}
\textbf{Nishiguchi, A., Yabe, T., Haines, et al. 1984, Phys. Rev. Lett., 53, 262}

\bibitem[Willingale et al. (2010a)]{Willingale10a}
\textbf{Willingale, L., Thomas, A. G. R., Nilson, P. M., et al. 2010, Phys. Rev. Lett., 105, 095001}

\bibitem[Willingale et al. (2010b)]{Willingale10b}
\textbf{Willingale, L., Nilson, P. M., Kaluza, M. C., et al. 2010, Phys. Plasmas, 17, 043104}

\bibitem[Gibbson (2005)]{Gibbson05}
\textbf{Gibbson, P. 2005, Phys. Rev. E, 72, 026411}

\bibitem[Kugland et al. (2012)]{Kugland12}
Kugland, N. L., Ryutov, D., Chang, P.-Y., et al. 2012, Nature Physics, 8, 809

\bibitem[Park et al. (2012)]{Park12}
Park, H.-S., Ryutov, D., Ross, J. S., et al. 2012, High Energy Density Physics, 8, 38

\bibitem[Ross et al. (2012)]{Ross12}
Ross, J. S., Glenzer, S. H., Amendt, P., et al. 2012, Phys. Plasmas, 19, 056501


\end{thebibliography}
\end{document}